\documentclass{ieeeaccess}
\usepackage{cite}
\usepackage{amsmath,amssymb,amsfonts}

\usepackage{graphicx}
\usepackage{textcomp}
\usepackage{caption}
\usepackage{CJK}
\usepackage{indentfirst}
\allowdisplaybreaks[4]
\usepackage{url}
\usepackage{subfigure}
\usepackage{algorithm}
\usepackage{algorithmic} 

\usepackage{amsmath}
\usepackage{graphics}
\usepackage{color,soul}
\usepackage{multicol}
\usepackage{bbding}

\usepackage{graphicx}

\def\BibTeX{{\rm B\kern-.05em{\sc i\kern-.025em b}\kern-.08em
    T\kern-.1667em\lower.7ex\hbox{E}\kern-.125emX}}
\begin{document}
\history{Date of publication xxxx 00, 0000, date of current version xxxx 00, 0000.}
\doi{10.1109/ACCESS.2017.DOI}

\title{LSTM-based Anomaly Detection for Non-linear Dynamical System}
\author{\uppercase{Yue Tan}\authorrefmark{1}, 
	\uppercase{Chunjing Hu}\authorrefmark{2}, 
	\uppercase{Kuan Zhang}\authorrefmark{3} \IEEEmembership{Member, IEEE},
	\uppercase{Kan Zheng}\authorrefmark{1} \IEEEmembership{Senior Member, IEEE},
	\uppercase{Ethan A. Davis}\authorrefmark{4},
	and \uppercase{Jae Sung Park}\authorrefmark{4}}

\address[1]{Intelligent Computing and Communication (ICC) Lab, Beijing University of Posts and Telecommunications, Beijing, China}
\address[2]{Key Laboratory of Universal Wireless Communications, Beijing University of Posts and Telecommunications, Beijing, China}
\address[3]{Department of Electrical and Computer Engineering, University of Nebraska-Lincoln, USA}
\address[4]{Department of Mechanical and Materials Engineering, University of Nebraska-Lincoln, USA}

\tfootnote{This work was supported by the National Natural Science Foundation of China (No. xxxx).}

\markboth
{Yue Tan \headeretal: Preparation of Papers for IEEE TRANSACTIONS and JOURNALS}
{Yue Tan \headeretal: Preparation of Papers for IEEE TRANSACTIONS and JOURNALS}

\corresp{Corresponding author: Chunjing Hu (e-mail: hucj@bupt.edu.cn).}

\begin{abstract}
Anomaly detection for non-linear dynamical system plays an important role in ensuring the system stability. However, it is usually complex and has to be solved by large-scale simulation which requires extensive computing resources. In this paper, we propose a novel anomaly detection scheme in non-linear dynamical system based on Long Short-Term Memory (LSTM) to capture complex temporal changes of the time sequence and make multi-step predictions. Specifically, we first present the framework of LSTM-based anomaly detection in non-linear dynamical system, including data preprocessing, multi-step prediction and anomaly detection. According to the prediction requirement, two types of training modes are explored in multi-step prediction, where samples in a wall shear stress dataset are collected by an adaptive sliding window. On the basis of the multi-step prediction result, a Local Average with Adaptive Parameters (LAAP) algorithm is proposed to extract local numerical features of the time sequence and estimate the upcoming anomaly. The experimental results show that our proposed multi-step prediction method can achieve a higher prediction accuracy than traditional method in wall shear stress dataset, and the LAAP algorithm performs better than the absolute value-based method in anomaly detection task.
\end{abstract}

\begin{keywords}
LSTM, anomaly detection, non-linear dynamical system, multi-step prediction, time series
\end{keywords}

\titlepgskip=-15pt

\maketitle

\section{Introduction}
\label{introduction}
\PARstart{D}{ynamical} system is the basic framework for modeling and control of an enormous variety of complicated systems, including fluid dynamics, signal propagation and interference in electronic circuits, heat transfer, biological systems, chemically reacting flows, etc. \cite{benner2015survey}. Nonlinearity is an important feature of these complex dynamical systems, as a result of a rich diversity of observed dynamical behaviors across the physical, biological, and engineering sciences. Modern non-linear dynamical systems are becoming more and more complex, where a variety of complicated phenomena make them vulnerable to software and hardware problems, causing anomalies in various emerging applications. It is necessary to predict and recover anomalies in time. However, in practice, this task is still manually solved. It is essential to provide automatic anomaly prediction methods for non-linear dynamical systems.\par

As time goes by, a large amount of time series data are produced by non-linear dynamical system. Some of them can be obtained by non-linear mapping and derivation of differential equations \cite{maulik2019subgrid}. During the past few decades, some techniques, such as Large-Eddy Simulation (LES) and Reynolds-averaged Navier–Stokes (RANS), were proposed to predict turbulent flow in grid-resolved scales accurately. However, LES requires a dedicated model for the effect on grid-resolved quantities \cite{meneveau2000scale, moeng1984large}. RANS also need to model the turbulence first in a temporally averaged sense \cite{bassi2005discontinuous, mortensen2011fenics}. To raise the advance anomaly alerts, non-linear dynamical systems should be continuously monitored.\par

Although the non-linear governing equations are usually known, simulations have to take extended periods of time and become computationally expensive, and time resolution with a certain accuracy is needed \cite{ takeishi2017subspace}. Deep learning is a useful method for modeling analysis and achieves a good result in many tasks including video classification, speech recognition, and natural language processing \cite{graves2008novel, yue2015beyond}. Deep learning methods also obtain results with high accuracy in complex prediction problems. However, there are few results based on deep learning methods in the field of non-linear dynamical system modeling.\par

We propose an LSTM-based method to the anomaly prediction task for non-linear dynamical system. Compared with traditional prediction methods, the LSTM-based method achieves a higher prediction accuracy for different zones of the time series. In particular, by incorporating LSTM into the developed anomaly detection algorithm, the temporal features of time series data are extracted in order to predict multi-step wall shear stress value of non-linear system. The main contributions of this paper lie in the following aspects.\par

\begin{itemize}
	\item We identify the problem of predicting multi-step wall shear stress and conduct experiments in a non-linear dynamical system. An LSTM-based anomaly detection method is proposed to solve this prediction problem. The proposed scheme employs a multi-layer network based on LSTM units, which captures complex temporal influences and pickup-drop-off interactions effectively.
	\item The anomaly points that reflect the latent danger to the system are identified. An effective anomaly detection algorithm, named Local Average with Adaptive Parameters (LAAP), is developed to exploit the anomalous period on predicted data. This algorithm is easily incorporated into the prediction model, to boost the performance of anomaly detection.
	\item Extensive experiments are conducted to evaluate the performance of the proposed methods. The dataset contains time series value of wall shear stress collected from a fluid non-linear dynamical system. The results show that our method achieves a higher prediction accuracy in both multi-step prediction and anomaly detection than a typical inference algorithm called Autoregressive Integrated Moving Average (ARIMA). 
\end{itemize}

The remainder of this paper is organized as follows. Section \ref{RW} introduces the related work on LSTM-based time series prediction and anomaly detection. Section \ref{method} presents a detailed description of the proposed prediction methods and anomaly detection algorithm. In Section \ref{experiment}, the experimental results and analysis depending on the proposed approach are presented. Finally, conclusions are given in Section \ref{conclusion}.\par

\section{Related Work}
\label{RW}
Time series prediction and anomaly detection have been widely studied and applied to a variety of real-world projects. In this section, related works on anomaly detection are reviewed. Related approaches on neural network for prediction are discussed.\par

\subsection{Anomaly detection methods}

Anomaly detection is a proactive method that raises alerts when the system is still in the normal state but progressing to an anomaly state \cite{chandola2009anomaly}. To prevent the anomaly or reduce the damage, anomaly detection is of great significance in many fields, such as geology, meteorology, and medicine.\par

For anomaly detection, the Markov model-based approaches are widely used. A Markov model is a stochastic model used to model randomly changing systems. It can be used to predict the state in the future using previous information. It is able to establish the transition probabilities relationship between states. In Markov chain model, the pattern of different metric values can be recognized and the next state of these values is predicted by the model. For example, Gu \textit{et al.} presented a stream-based mining algorithm by combining Markov models and Bayesian classification methods for online anomaly prediction. Their anomaly prediction scheme can raise alerts for impending system anomalies in advance and suggests possible anomaly causes \cite{gu2009online}. Sendi \textit{et al.} proposed a framework to predict multi-step attacks before they pose a serious security risk. They used Hidden Markov Model (HMM) to extract the interactions between attackers and networks \cite{sendi2012real}. Paulo \textit{et al.} applied a Markov chains-based method to characterize the stochasticity of droughts and predict the transition probability from one class of severity to another up to 3 months ahead \cite{paulo2007prediction}. In \cite{zhou2013novel}, Zhou \textit{et al.} employed the Evidential Markov chain for the anomaly prediction of PlanetLab. A Belief Markov chain is proposed to extend the Evidential Markov chain and cope with noisy data stream.\par

\begin{figure*}[!h]
	\centering
	\includegraphics[width=0.9\textwidth]{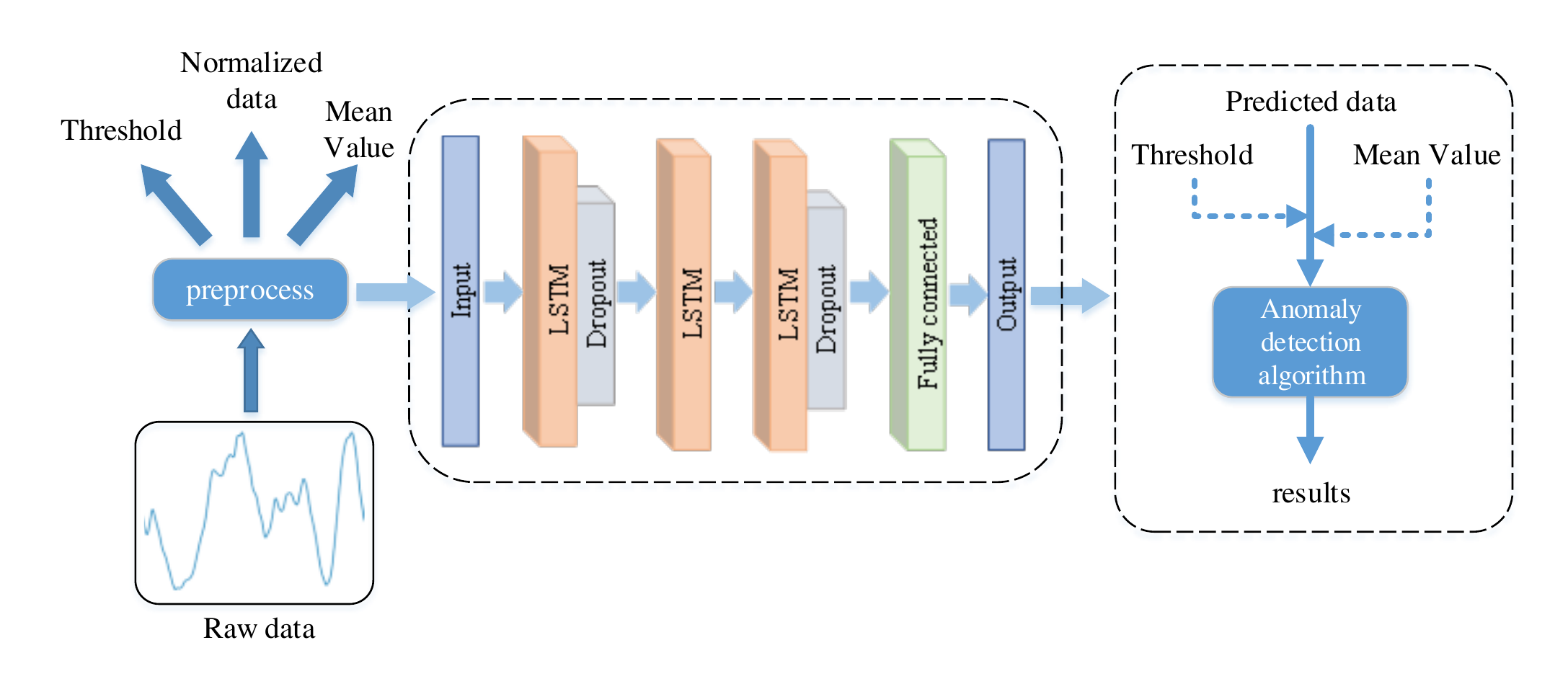}
	\caption{The framework of LSTM-based anomaly detection approach}
	\label{pipeline}
\end{figure*}

Also, there are many anomaly detection approaches based on regression methods. In these approaches, the detection problems are converted into normal regression problems, then machine learning-based regression algorithms and models can be applied to anomaly detection. For instance, Hong et.al proposed a new anomaly detection approach based on principal component analysis (PCA), information entropy theory and support vector regression (SVR). It can be used in credit card fraud detection as well as intrusion detection in cyber-security \cite{hong2016entropy}. In \cite{huang2016using}, Huang \textit{et al.} presented a Recurrent Neural Networks (RNN) model for the anomaly prediction of component-based enterprise systems. Their RNN-based method has shown high prediction accuracy and time efficiency for large-scale systems.\par

\subsection{LSTM for prediction}

In recent years, neural network-based deep learning approaches are widely applied to the prediction problems \cite{chandola2009anomaly}. Among them, Recurrent Neural Network (RNN) and its advanced variants have shown a higher performance for prediction tasks than traditional methods. RNN is a type of artificial neural networks. It can capture the feature of input time series data by remembering its historical information. LSTM Network is a special RNN. LSTM is proved to outperform many other types of RNN in modeling sequential data and widely used in prediction tasks \cite{peng2017deep}.\par

RNN is suitable for learning patterns, relationships and interconnections hidden in time series as well as modeling the temporal sequences. In \cite{zio2012failure}, Zio \textit{et al.} employed Infinite Impulse Response Locally Recurrent Neural Networks (IIR-LRNN) to forecast failures and make reliability prediction for systems' components. It is innovative to use such dynamic modeling technique in reliability prediction tasks.\par 

Especially, LSTM, which was proposed by Hochreiter \textit{et al.} in 1997 \cite{hochreiter1997long}, has emerged to be an effective and scalable type of RNNs for several learning problems related to sequential data \cite{greff2016lstm}. By utilizing multiplicative gates that enforce constant error flow through the internal states of cells, LSTMs overcome the vanishing gradient problem in original RNNs \cite{malhotra2015long}. 

Due to their superior ability for processing time-series data, LSTMs are widely applied to prediction tasks. In \cite{duan2016travel}, an LSTM-based deep learning method is explored for travel time prediction on the dataset provided by Highways England. They have achieved high prediction accuracy for 1-step ahead travel time prediction error. In \cite{jia2017incremental}, an LSTM-based spatio-temporal learning framework is proposed for land cover prediction. The authors design a dual-memory structure to capture both long-term and short-term patterns in temporal sequences. Based on LSTM, the authors in \cite{yuan2018incorporating} proposed an improved model to learn tweet representation from weakly-labeled data and make tweet classification with higher accuracy. In \cite{peng2017deep}, LSTM network is first used to predict the performance of web servers as the URL requests are always sequential data. The logs of Nginx web servers are analyzed before the performance of web server is predicted.

Besides, Chen \textit{et al.} utilized LSTM for the prediction task for China stock \cite{chen2015lstm}.  Altche \textit{et al.} addressed a highway trajectory prediction approach based on LSTM network \cite{altche2017lstm}. Qu \textit{et al.} applied an approach based on PCA and LSTM to the field of wind power prediction \cite{xiaoyun2016short}.

\section{LSTM-Based Anomaly Detection}
\label{method}

In this section, LSTM-based multi-step prediction for non-linear dynamic system is presented in detail. Then, we propose an adaptive anomaly detection method and a local average algorithm with adaptive parameters.\par


In our proposed method, the first and last LSTM layers are followed by a dropout layer, which helps prevent overfitting. The last dropout layer is followed by a fully connected layer, as shown in Fig. \ref{pipeline}. In the framework of the LSTM-based anomaly detection approach, the first segment is the data preprocessing procedure. After preprocessing, the mean value and threshold are calculated for the following parts. The raw data are transformed into normalized data. The second segment is the network architecture for multi-step prediction. It is composed of three kinds of network layers which include LSTM layer, dropout layer, and a fully connected layer. The seven-layer network outputs multi-step prediction results of the future time series values. In the third segment, the prediction results produced in the second segment are used as input and anomaly detection results are generated according to the local average algorithm with adaptive parameters.\par

\subsection{LSTM-Based Multi-step Prediction}

\begin{figure}[!t]
	\centering
	\subfigure[single timestep]{
		\begin{minipage}[b]{0.22\textwidth}
			\includegraphics[width=1\textwidth]{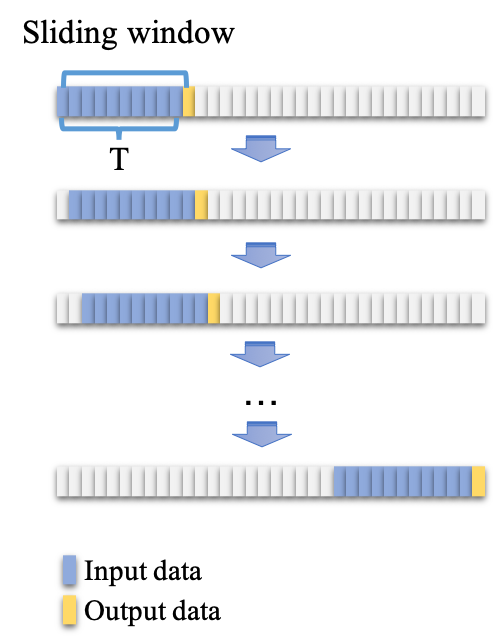}
		\end{minipage}		
		\label{mode1}
	}
	\subfigure[multi-timestep]{
		\begin{minipage}[b]{0.22\textwidth}
			\includegraphics[width=1\textwidth]{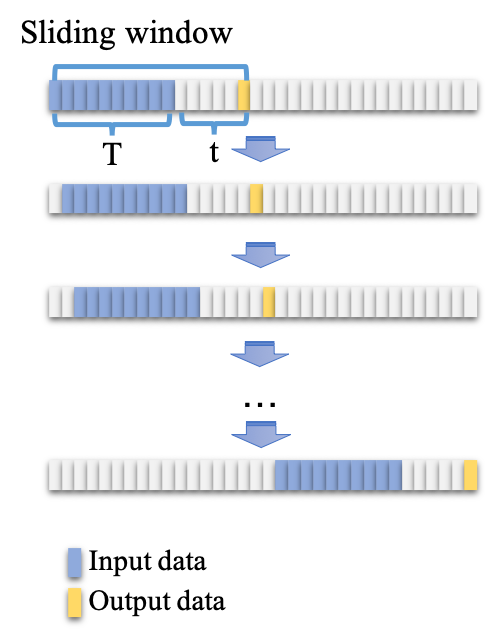}
		\end{minipage}		
		\label{mode2}		
	} 
	\caption{Two types of training modes}
	\label{modes}
\end{figure}

LSTM layer receives a time sequence of the same length $N$ as input and outputs them to dropout layer to prevent overfitting. To collect samples for training, there are two different modes according to the requirement of prediction demand. As is shown in Fig. \ref{modes}, there are two types of training modes according to the requirement of prediction demand. $T$ refers to the length of the input time series and $t$ means the prediction of the $t^{th}$ point after the current point. 

In mode $a$, only the next point of current input time sequence is predicted in each prediction, as is illustrated in Fig. \ref{mode1}. A sample used for training is obtained by current state window. It contains the time sequence of fixed length and the value of the next timestep. By sliding over the long time series used for training, all samples are gathered into the training set. Mode $b$, as shown in Fig. \ref{mode2}, meets the requirement of muti-timestep prediction. For the sliding window in mode $b$, it equals to $T+t$ which contains the time sequence and a time interval between current value and the value to predict. In this mode, a training sample consists of time sequence of fixed length and the value of the $t^{th}$ timestep after the current value.

When all training samples are gathered into the training set, the training process begins. It randomly selects from the training set to remove correlations in the sequence and smooths the changes in data distribution. 

Irrespective of different training modes, the Root Mean Square Error (RMSE) is utlized to evaluate the prediction performance of the proposed method,

\begin{equation}
\operatorname{RMSE}=\sqrt{\frac{1}{Q-T-t+1}} \sum_{i=1}^{Q-T-t+1}\left(y_{i}-\hat{y}_{i}\right)^{2},
\end{equation}

\noindent where $y_i$ and $\hat{y}_{i}$ are the predicted value and the ground truth for timestep $i$, respectively, and $Q$ is the total length of non-linear dynamical dataset used for training. Traditional models that be used for multi-timestep prediction include Auto Regressive Integrated Moving Average (ARIMA), eXtreme Gradient Boosting (XGBoost), Logistic Regression (LR), etc. Among them, ARIMA is used as our baseline model for comparison in this paper.

\subsection{Adaptive Anomaly Detection}

Another important module in the architecture of LSTM-based anomaly detection method is the following anomaly detection. An algorithm, named Local Average with Adaptive Parameters (LAAP), is put forward. This algorithm uses adaptive parameters to compute local numerical features including average, standard deviation and slope. The threshold for determining anomaly depends on those features of predicted timestep.

\begin{algorithm}[t]
	\caption{Local Average with Adaptive Parameters}
	\hspace*{0.02in} {\bf Input:} 
	
	\hspace*{0.06in}the predicted time series, $D : s_{1}, s_{2}, \dots, s_{d}$
	
	\hspace*{0.06in}window length, $W$
	
	\hspace*{0.06in}parameter adaptive rate, $\alpha$
	
	\hspace*{0.02in} {\bf Process:} 
	
	\begin{algorithmic}[1]
		\STATE Compute the local average ${\mu}_i$ in a window of length $W$ for every point in the predicted time series by \eqref{avg}
		\STATE Compute the local standard deviation ${\sigma}_i$ in a window of length $W$ for every point in the predicted time series by \eqref{lstd}
		\STATE Compute the local slope $k_i$ in a window of length $W$ for every point in the predicted time series,
		$
		k_{i}=\frac{1}{W}\left(s\left(i+\frac{W}{2}\right)-s\left(i-\frac{W}{2}\right)\right)
		$
		\IF {$k(i)<0$}
		\STATE 
		$
		y_i=\left\{\begin{array}{ll}{1} & {\text { if } s_i<{\mu}_{i}-\alpha \cdot {\sigma}_{i}} \\ {0} & {\text { otherwise }}\end{array}\right.
		$
		\ELSE
		\STATE 
		$
		y_i=\left\{\begin{array}{ll}{1} & {\text { if } s_i>{\mu}_{i}+\alpha \cdot {\sigma}_{i}} \\ {0} & {\text { otherwise }}\end{array}\right.
		$
		\ENDIF
	\end{algorithmic}
	
	\hspace*{0.02in} {\bf Output:} $y_i$
	\label{alg}
\end{algorithm}

As the sliding window moves forward, the predicted time series can be obtained. The length of the predicted future timesteps is assumed to be $d$. The predicted time series can be denoted as $D : s_{1}, s_{2}, \dots, s_{d}$. The window length and parameter adaptive rate are set to be $W$ and $\alpha$ respectively. The value of $W$ is usually determined by the pattern of most anomaly. The parameter adaptive rate can be adjusted according to how sensitive the detection system should be. It ranges from 0 to 1. If the parameter adaptive rate is relatively high, the anomaly detection system is more sensitive to the upcoming anomaly. Otherwise, the system may reduce the probability for predicting an upcoming anomaly.

In each window length of $W$, the local average ${\mu}_i$ is computed as

\begin{equation}
{\mu}_i=\frac{1}{W} \sum_{l=i-\frac{W}{2}}^{i+\frac{W}{2}} s_l,
\label{avg}
\end{equation}

\noindent and the local standard deviation ${\sigma}_i$ is computed as

\begin{equation}
{\sigma}_i=\sqrt{\frac{1}{W} \sum_{k=i-\frac{W}{2}}^{i+\frac{W}{2}}\left((s_k-{\mu}_i)^{2}\right)}.
\label{lstd}
\end{equation}

\noindent Then, a local slope $k_i$ is computed to discriminate different kinds of potential anomalies. Generally, there are two kinds of potential anomalies in the non-linear dynamical system. 

One kind of anomalies appears to be a local maximum while another kind of anomalies appears to be a local minimum. According to these two different situations, the result of anomaly detection $y_i$ is obtained and output by algorithm LAAP. To summarize the above procedure of our proposed method, the pseudocode of LAAP is shown in Algorithm \ref{alg}.

\section{Experimental results and analysis}
\label{experiment}
In this section, experiments on the basis of non-linear dynamical data are carried out to demonstrate the effectiveness of our proposed method. 

\subsection{Dataset}
To test and verify the LSTM-based multi-step prediction and the following anomaly detection, the data is generated by the temporal evolution of an incompressible Newtonian fluid in the plane Poiseuille (channel) geometry where the flow is driven by a constant volumetric flux, $Q$, at a Reynolds number of $R_e=1800$. The $x$, $y$, and $z$ coordinates are aligned in the streamwise, wall-normal, and spanwise directions, respectively. Periodic boundary conditions are imposed in the $x$ and $z$ direction with fundamental periods of $L_x$ and $L_z$. No-slip boundary conditions are imposed at the top and bottom walls $y=\pm h$, where $h=L_{y} / 2$ is the half-channel height. Such a boundary condition necessitates that streamwise velocity is zero at both walls. In this study, the so-called minimal flow unit methodology is used with a domain size of $L_{x} \times L_{y} \times L_{z}=2 \pi \times 2 \times \pi$ \cite{jimenez1991minimal}. A numerical grid system is generated on $N_{x} \times N_{y} \times N_{z}$ meshes, where Fourier-Chebyshev-Fourier spectral spatial discretization is applied to all variables. With a mesh convergence study, the resolution used is $\left(N_{x}, N_{y}, N_{z}\right)=(48,81,48)$. The time step used for forward integration of the system is $\Delta t=0.02$. Given these temporal and spatial resolutions, the corresponding errors are $O\left(10^{-6}\right)$ and $O\left(10^{-4}\right)$, respectively. For more details, the reader is referred to \cite{park2015exact, park2018bursting, whalley2017low}.\par

For this study, the one-dimensional data used for our experiments is the time series of the wall shear stress $\tau=\mu \partial u / \partial y$, where $\mu$ is the dynamic viscosity of the fluid and $u$ is the streamwise velocity. The wall shear stress is a measure of the resistance of fluid experiences and is a qualitative measure of the state of nonlinear dynamical system.\par

\begin{figure}
	\centering
	\includegraphics[width=0.5\textwidth]{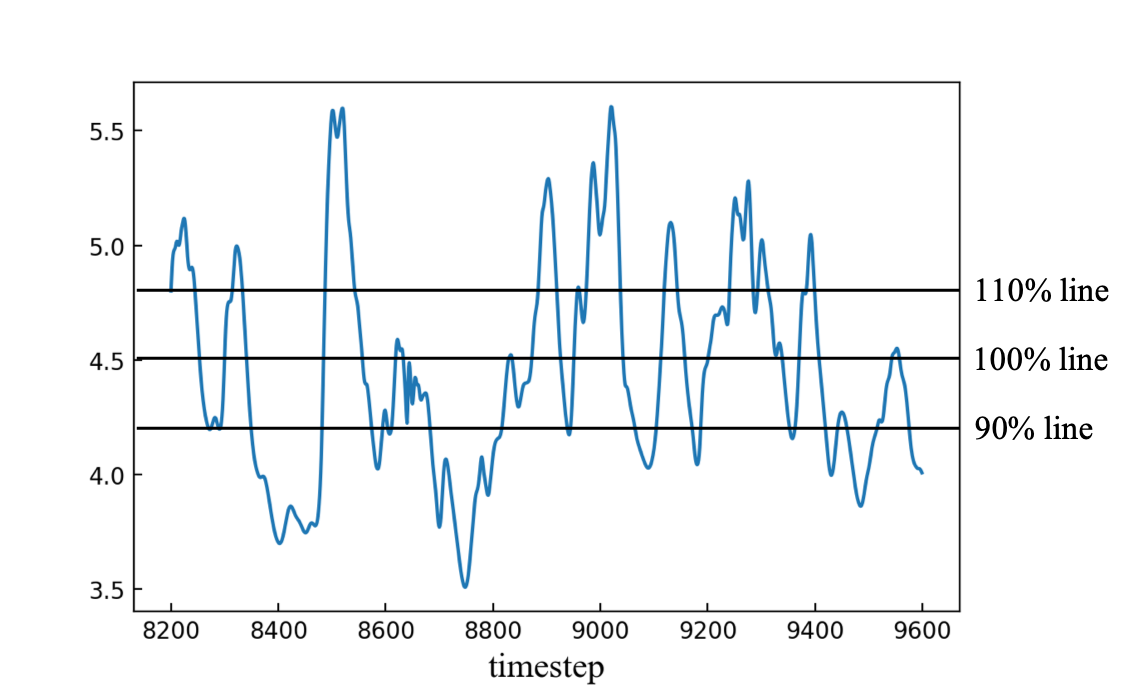}
	\caption{A segment of the raw data.}
	\label{rawdata}
\end{figure}

A segment of the raw data is visualized in Fig. \ref{rawdata}. The data fluctuates around a certain value and has a definite boundary in normal state. According to the value, time series can be divided into four zones: those below 90\% of the mean value (zone 1), those between 90\% of the mean value and the mean value (zone 2), those between the mean value and 110\% of the mean value (zone 3), and those over 110\% of the mean value (zone 4). Data in zone 4 should be recognized because there is a great probability of a turbulence. Usually, there is a high transition probability from zone 1 to zone 4, and anomalies are more likely to appear in zone 1 and zone 4. Hence, the main objective of the anomaly detection is to predict the data in zone 1 and zone 4. Accurate prediction of these anomalies can effectively prevent some potential accidents from happening.\par

\subsection{Results and Analysis}

In this part, the experimental results are given and analyzed from two aspects: multi-step prediction and anomaly detection.

\subsubsection{Multi-step prediction}

\begin{figure*}[!h]
	\centering
	\subfigure[predicted timestep = 4]{
		\begin{minipage}[b]{0.48\textwidth}
			\includegraphics[width=1\textwidth]{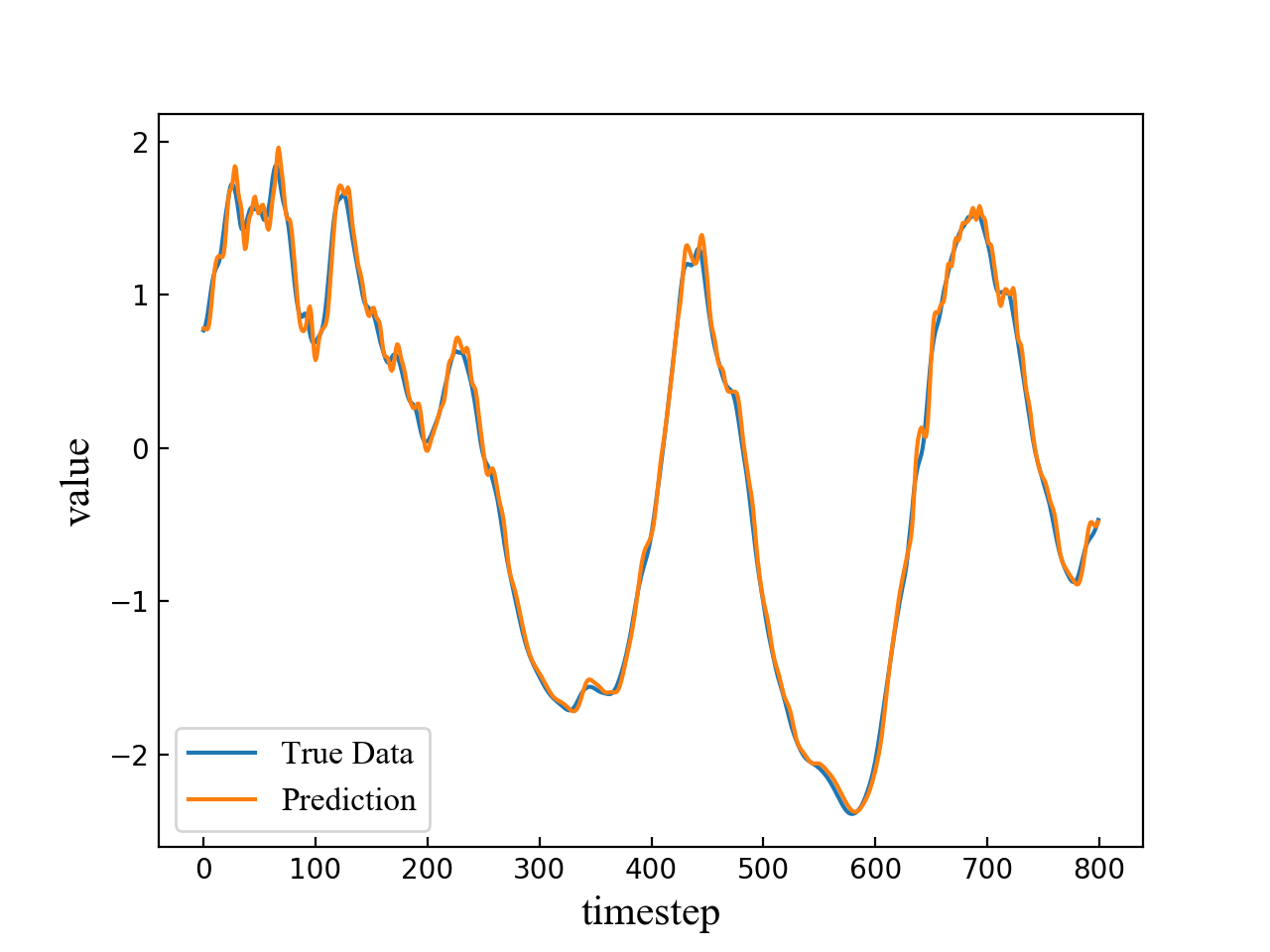}
		\end{minipage}		
		\label{res1}
	}
	\subfigure[predicted timestep = 6]{
		\begin{minipage}[b]{0.48\textwidth}
			\includegraphics[width=1\textwidth]{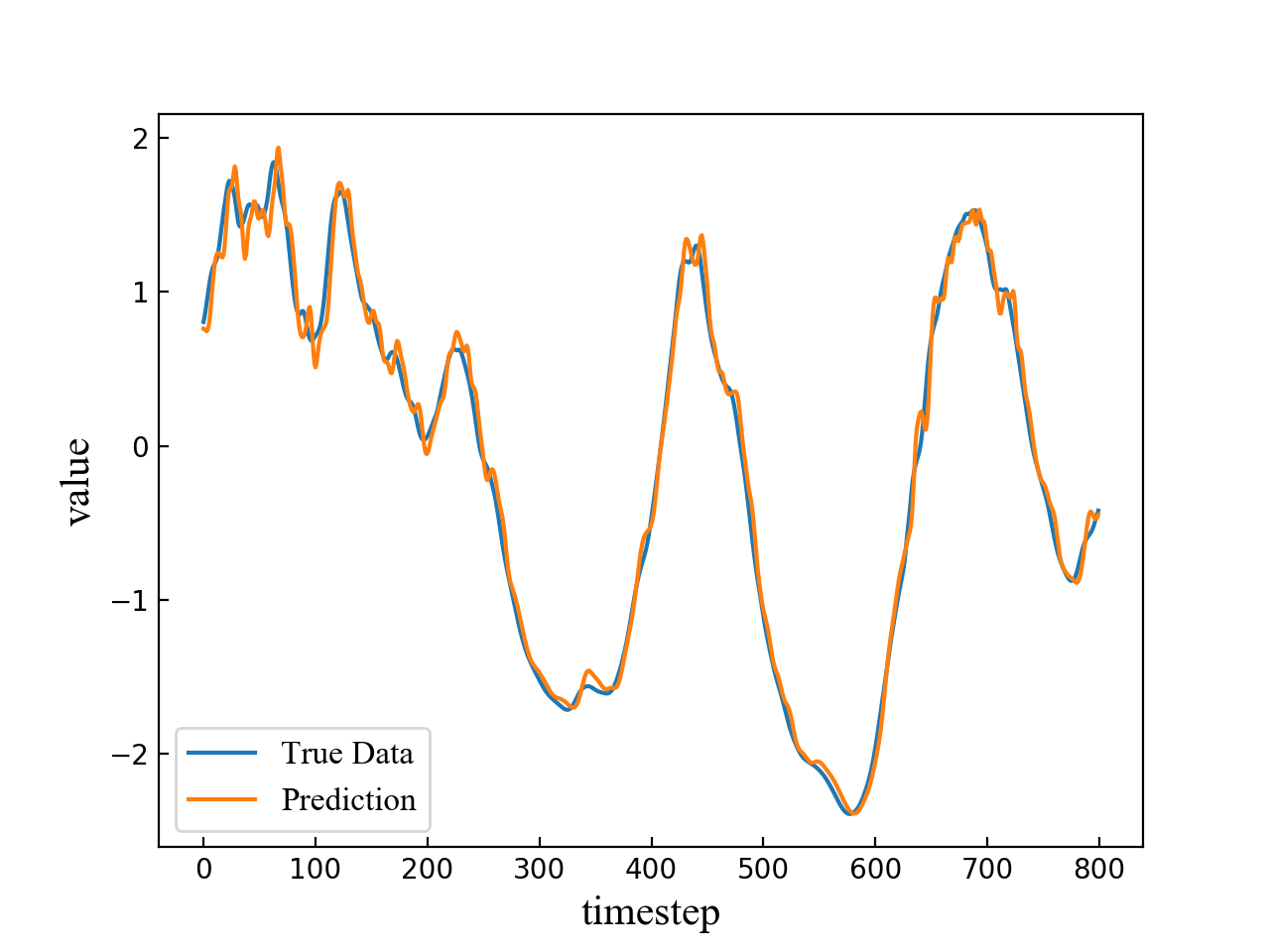}
		\end{minipage}		
		\label{res2}		
	}
	\subfigure[predicted timestep = 8]{
		\begin{minipage}[b]{0.48\textwidth}
			\includegraphics[width=1\textwidth]{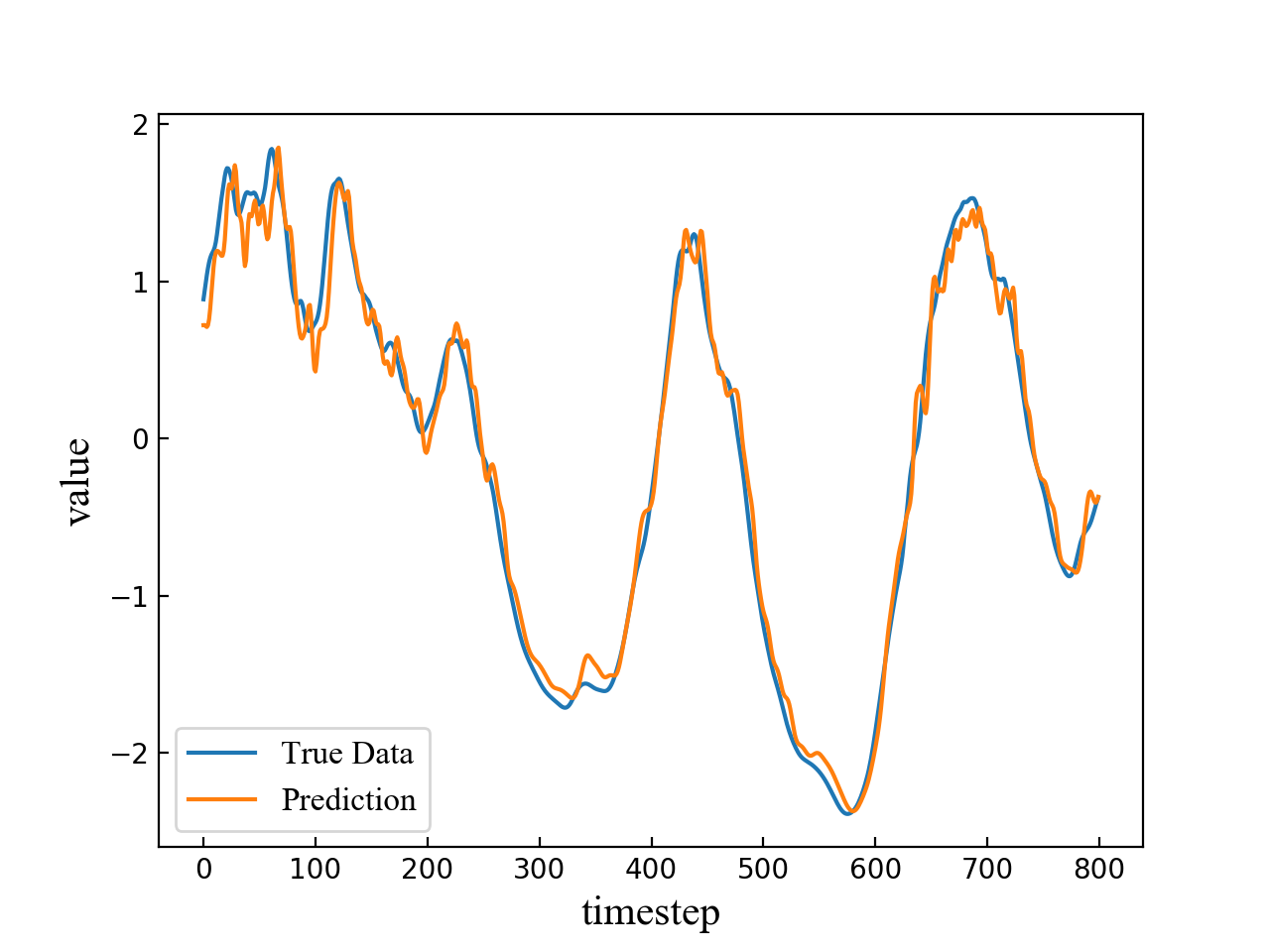}
		\end{minipage}		
		\label{res3}		
	}
	\subfigure[predicted timestep = 10]{
		\begin{minipage}[b]{0.48\textwidth}
			\includegraphics[width=1\textwidth]{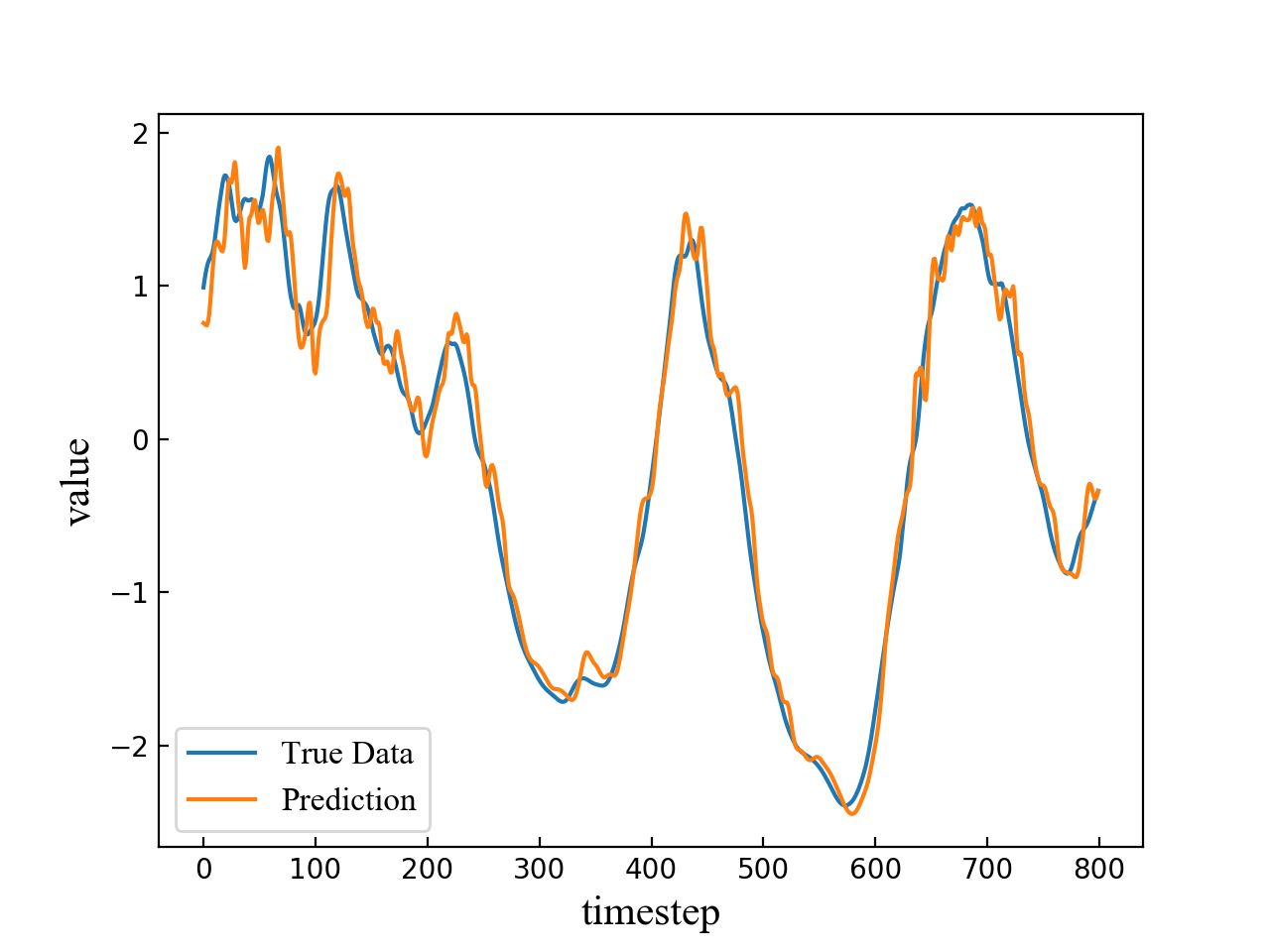}
		\end{minipage}		
		\label{res4}		
	}
	\caption{LSTM-based multi-step prediction result.}
	\label{predict_result}
\end{figure*}

The length of the input time sequence is 50 and the predicted timestep is set to be 4, 6, 8, 10 respectively. A segment of the test result consisting of 800 points is shown in Fig. \ref{predict_result}. The blue line refers to true data while the orange line refers to multi-step prediction result. As is shown in the four subfigures in Fig. \ref{predict_result}, the prediction error becomes larger and the fluctuation becomes more obvious when the predicted timestep increases.\par 

Different LSTM network models are trained to predict the points of different step sizes. The whole data are split into two parts, i.e., training set and test set. The ratio of these two sets is 0.8. In this way, about 4000 samples are used for training and 1000 samples for testing. Table 1 shows the precision of zone 1 and zone 4 prediction. The mean value of the data set is 4.042179. The upper threshold is 4.446397 and the lower threshold is 3.637961. LSTM has achieved a high precision rate in both zone 1 and zone 4 prediction. Table \ref{precisionofzone} shows the prediction accuracy of zone 1 and zone 4 as the prediction timestep ranges from 1 to 10.


\begin{table}[htb]
	\caption{Prediction accuracy of zone 1 and zone 4}
	\begin{center}
		\begin{tabular}{p{30pt} p{40pt} p{40pt} }
			\hline
			Timestep  &	Accuracy of zone 1	 &   Accuracy of zone 4  \\
			\hline
			1 & 0.9909 & 0.9991 \\
			2 & 0.9913 & 0.9932 \\
			3 & 0.9801 & 0.9949 \\
			4 & 0.9755 & 0.9910 \\
			5 & 0.9752 & 0.9850 \\
			6 & 0.9633 & 0.9807 \\
			7 & 0.9469 & 0.9850 \\
			8 & 0.9490 & 0.9743 \\
			9 & 0.9357 & 0.9739 \\
			10 & 0.9346 & 0.9713\\
			
			\hline
			
		\end{tabular}
		\label{precisionofzone}
	\end{center}
\end{table}

The prediction result of the LSTM-based method is compared with the ARIMA. When the predicted step size is larger than 2, LSTM performs better than ARIMA as the RMSE of LSTM predicted results is lower than that of the ARIMA. Furthermore, LSTM shows a stable growth, which means that different requirements of accuracy can be met by adjusting the predicted step size. The parameters for the ARIMA model $ar$ and $ma$ are both set to be 2 in our experiment. Then, the maximum likelihood estimation is used to fit our return rate to the ARIMA model. However, it should be noted that there are some irregular points in the ARIMA model. That is because the ARIMA model relies on the stability of temporal data and more stable data leads to better performance in prediction. However, the prediction of dynamical system contains one or more unstable factors. The ARIMA cannot capture such a change in time like LSTM.

\begin{figure}
	\centering
	\includegraphics[width=0.48\textwidth]{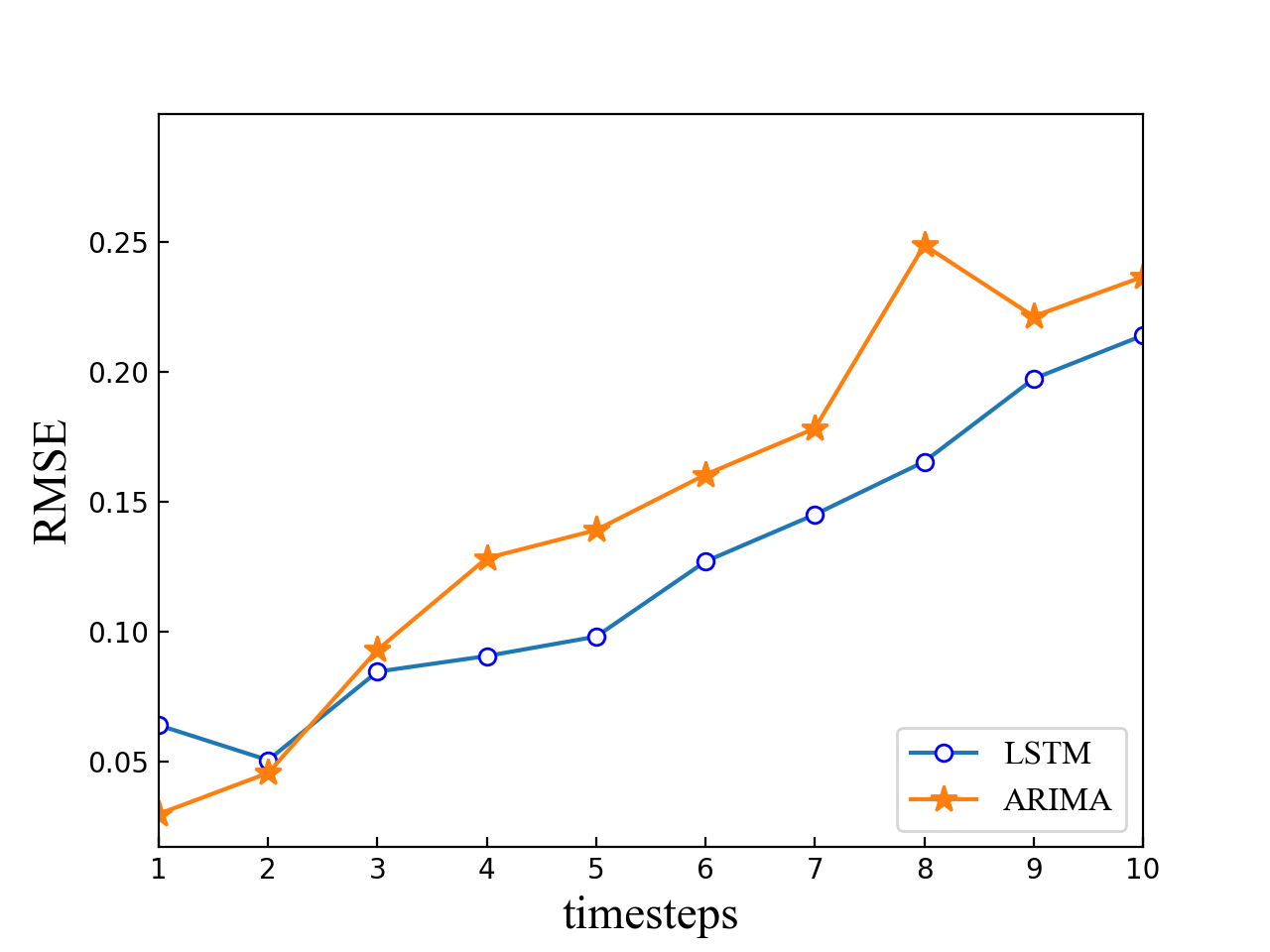}
	\caption{RMSE of different predicted step size}
	\label{RMSE}
\end{figure}

\subsubsection{Anomaly detection}

We aim to raise alerts before anomalies occur. So, the anomaly detection results are acquired based on the multi-step prediction results. In order to prevent anomalies in advance, the anomaly detection model should be efficient and accurate. In comparison with absolute value-based anomaly detection, the accuracy of our proposed algorithm LAAP is evaluated by an error distribution function.\par 

\begin{figure}[!h]
	\centering
	\subfigure[error distribution of absolute value-based anomaly detection algorithm]{
		\begin{minipage}[b]{0.48\textwidth}
			\includegraphics[width=1\textwidth]{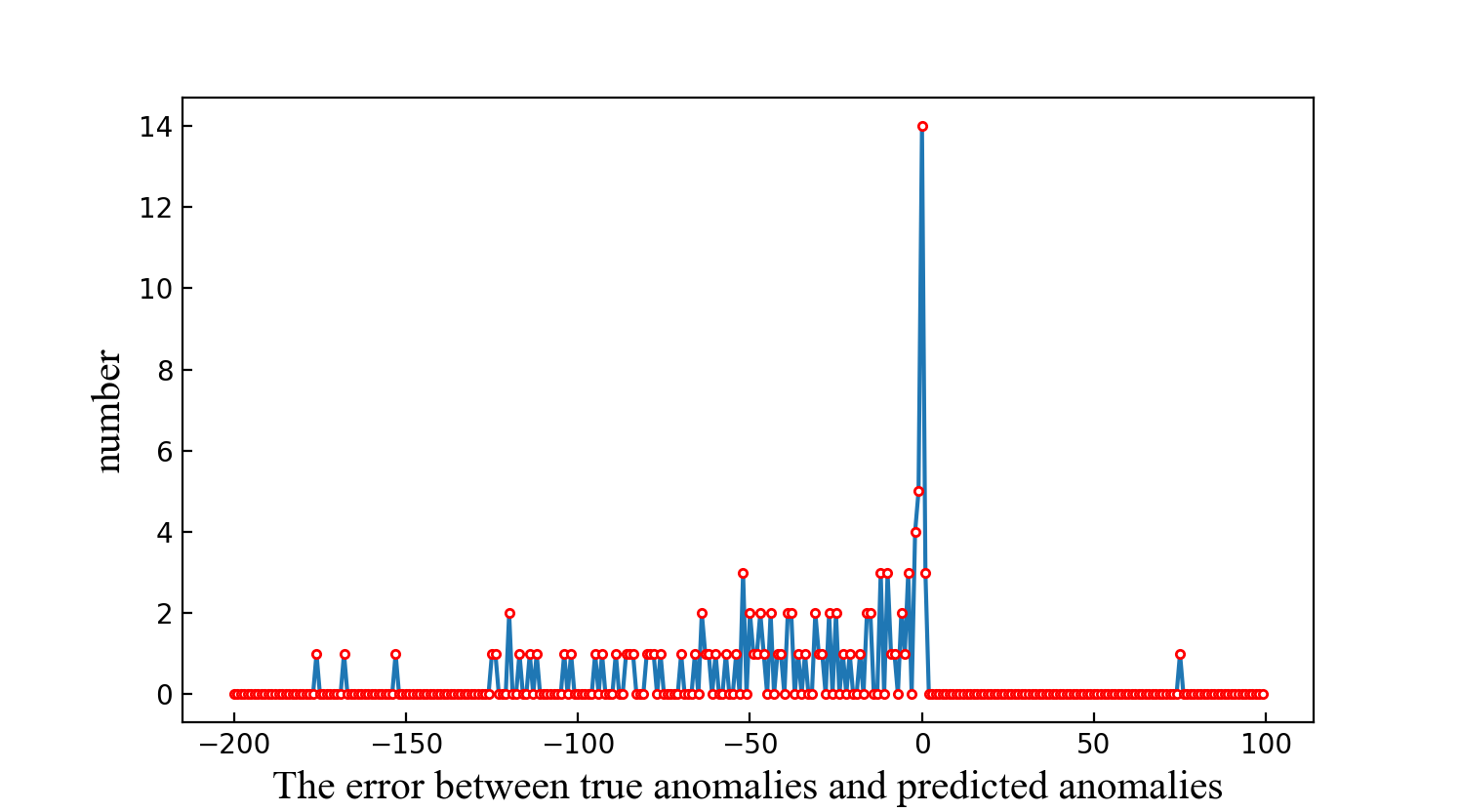}
		\end{minipage}
		\label{err_m1}
	}
	\subfigure[error distribution of LAAP algorithm]{
		\begin{minipage}[b]{0.48\textwidth}
			\includegraphics[width=1\textwidth]{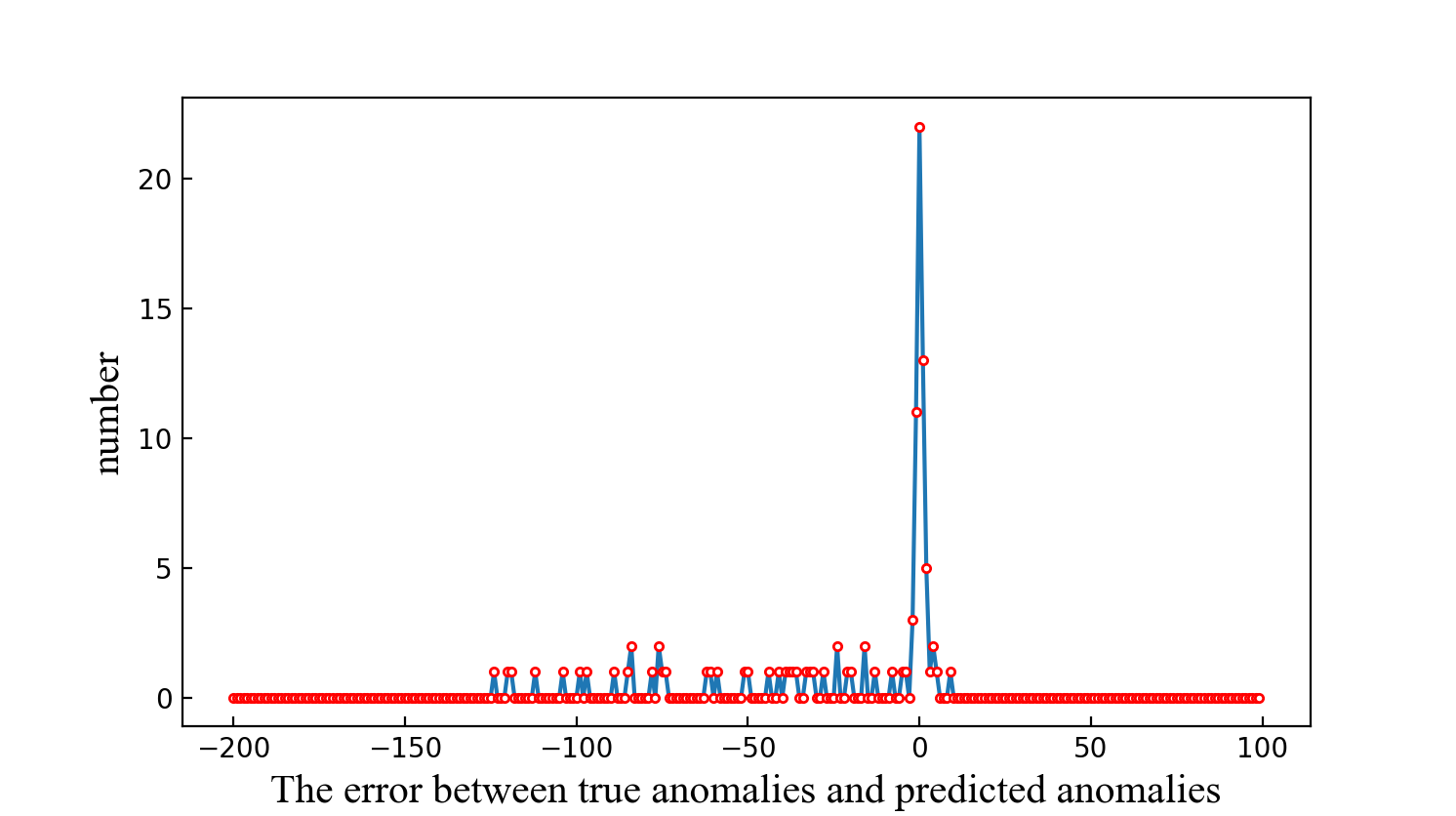}
		\end{minipage}
		\label{err_m2}
	} 
	\caption{Error distribution of anomaly detection Algorithm}
	\label{err_alg}
\end{figure}

Fig. \ref{err_m1} and Fig. \ref{err_m2} illustrate the error distribution of absolute value-based anomaly detection algorithm and the LAAP algorithm, respectively. The horizontal axis represents the time interval error between the nearest true anomaly and predicted anomaly and the vertical axis refers to the total amount of anomalies of a certain predictive error. In Fig. \ref{err_m1}, zero error occurs 14 times, and the time interval error ranges from around -175 to 75. While in Fig. \ref{err_m2}, zero error occurs more than 20 times, and the time interval error ranges from around -130 to 5. \par 

According to the error distribution results, the frequency of zero error by LAAP algorithm is larger than that computed by the absolute value-based anomaly detection algorithm. Moreover, the error range of LAAP is smaller than the absolute value-based anomaly detection algorithm. Therefore, it is more likely to accurately predict when an anomaly occurs by LAAP algorithm, and it can be concluded that LAAP performs better than the absolute value-based anomaly detection algorithm. \par

\section{Conclusion}
\label{conclusion}

In this paper, we have proposed an LSTM-based anomaly detection method for non-linear dynamical system. A sliding window scheme is used to collect training samples for multi-step prediction. Then, an LAAP algorithm is developed to make anomaly detection based on the LSTM prediction result. Experiments have been conducted to evaluate the performance of the proposed methods. The results indicate that our proposed multi-step prediction has achieved lower RMSE and higher prediction accuracy compared with ARIMA on wall shear stress dataset, and the LAAP algorithm outperforms the absolute value-based anomaly detection algorithm.

\bibliography{predict}
\bibliographystyle{IEEEtran}

%
%
%

\EOD

\end{document}